# QEMPAR: QoS and Energy Aware Multi-Path Routing Algorithm for Real-Time Applications in Wireless Sensor Networks

Saeed Rasouli Heikalabad[1], Hossein Rasouli[2], Farhad Nematy[2] and Naeim Rahmani[2]

[1] Technical and Engineering Department, Islamic Azad University - Tabriz Branch
Tabriz, East Azerbaijan, Iran

[2] Technical and Engineering Department, Islamic Azad University - Tabriz Branch
Tabriz, East Azerbaijan, Iran

**Abstract**
Enabling real time applications in wireless sensor networks requires certain delay and bandwidth which pose more challenges in the design of routing protocols. The algorithm that is used for packet routing in such applications should be able to establish a tradeoff between end to end delay parameter and energy consumption. In this paper, we propose a new multi path routing algorithm for real time applications in wireless sensor networks namely QEMPAR which is QoS aware and can increase the network lifetime. Simulation results show that the proposed algorithm is more efficient than previous algorithms in providing quality of service requirements of real-time applications.

## 1. Introduction

In the recent years, the rapid advances in micro-electro-mechanical systems, low power and highly integrated digital electronics, small scale energy supplies, tiny microprocessors, and low power radio technologies have created low power, low cost and multifunctional wireless sensor devices, which can observe and react to changes in physical phenomena of their environments. These sensor devices are equipped with a small battery, a tiny microprocessor, a radio transceiver, and a set of transducers that used to gathering information that report the changes in the environment of the sensor node. The emergence of these low cost and small size wireless sensor devices has motivated intensive research in the last decade addressing the potential of collaboration among sensors in data gathering and processing, which led to the creation of Wireless Sensor Networks (WSNs).

A typical WSN consists of a number of sensor devices that collaborate with each other to accomplish a common task (e.g. environment monitoring, target tracking, etc) and report the collected data through wireless interface to a base station or sink node. The areas of applications of WSNs vary from civil, healthcare and environmental to military. Examples of applications include target tracking in battlefields [1], habitat monitoring [2], civil structure monitoring [3], forest fire detection [4], and factory maintenance [5].

However, with the specific consideration of the unique properties of sensor networks such limited power, stringent bandwidth, dynamic topology (due to nodes failures or even physical mobility), high network density and large scale deployments have caused many challenges in the design and management of sensor networks. These challenges have demanded energy awareness and robust protocol designs at all layers of the networking protocol stack [6].

Efficient utilization of sensor's energy resources and maximizing the network lifetime were and still are the main design considerations for the most proposed protocols and algorithms for sensor networks and have dominated most of the research in this area. The concepts of latency, throughput and packet loss have not yet gained a great focus from the research community. However, depending on the type of application, the generated sensory data normally have different attributes, where it may contain delay sensitive and reliability demanding data. For example, the data generated by a sensor network that monitors the temperature in a normal weather monitoring station are not required to be received by the sink node within certain time limits. On the other hand, for a sensor network that used for fire detection in a forest, any sensed data that carries an indication of a fire should be reported to the processing center within certain time limits. Furthermore, the introduction of multimedia sensor networks along with the increasing interest in real time applications have made strict constraints on both throughput and delay in order to report the time-critical data to the sink within certain time limits and bandwidth requirements without any loss. These performance metrics (i.e. delay, energy consumption and bandwidth) are usually referred to as Quality of Service (QoS) requirements [7]. Therefore, enabling many applications in sensor networks requires energy and QoS awareness in different layers of the protocol stack in order to have efficient utilization of the network resources and effective access to sensors readings. Thus QoS routing is an important topic in sensor networks research, and it has been under the focus of the research community of WSNs. Authors of [7] and [8] have surveyed the QoS based routing protocol in WSNs.

Many routing mechanisms specifically designed for WSNs have been proposed [9][10]. In these works, the unique properties of the WSNs have been taken into account. These routing techniques can be classified according to the protocol operation into negotiation based, query based, QoS based, and multi-path based. The negotiation based protocols have the objective to eliminate the redundant data by include high level data descriptors in the message exchange. In query based protocols, the sink node initiates the communication by broadcasting a query for data over the network. The QoS based protocols allow sensor nodes to make a tradeoff between the energy consumption and some QoS metrics before delivering the data to the sink node [11]. Finally, multi-path routing protocols use multiple paths rather than a single path in order to improve the network performance in terms of reliability and robustness. Multi-path routing establishes multiple paths between the source-destination pair. Multi-path routing protocols have been discussed in the literature for several years now [12]. Mutli-path routing has focused on the use of multiple paths primarily for load balancing, fault tolerance, bandwidth aggregation, and reduced delay. We focus to guarantee the required quality of service through multi-path routing.

The rest of the paper organized as follows: in section 2, we explain the related works. Section 3 describes the proposed algorithm with detailed. Section 4 explore the simulation parameters and result analysis. Final section is containing of conclusion and future works.

## 2. Related Works

QoS-based routing in sensor networks is a challenging problem because of the scarce resources of a sensor node. Thus, this problem has received a significant attention from the research community, where many works are being made. Some QoS oriented routing works are surveyed in [7] and [8]. In this section we do not give a comprehensive summary of the related work, instead we present and discuss some works related to proposed protocol.

One of the early proposed routing protocols that provide some QoS is the Sequential Assignment Routing (SAR) protocol [13]. SAR protocol is a multi-path routing protocol that makes routing decisions based on three factors: energy resources, QoS on each path, and packet's priority level. Multiple paths are created by building a tree rooted at the source to the destination. During construction of paths those nodes which have low QoS and low residual energy are avoided. Upon the construction of the tree, most of the nodes will belong to multiple paths. To transmit data to sink, SAR computes a weighted QoS metric as a product of the additive QoS metric and a weighted coefficient associated with the priority level of the packet to select a path. Employing multiple paths increases fault tolerance, but SAR protocol suffers from the overhead of maintaining routing tables and QoS metrics at each sensor node.

K. Akkaya and M. Younis in [14] proposed a cluster based QoS aware routing protocol that employs a queuing model to handle both real-time and non real time traffic. The protocol only considers the end-to-end delay. The protocol associates a cost function with each link and uses the K-least-cost path algorithm to find a set of the best candidate routes. Each of the routes is checked against the end-to-end constraints and the route that satisfies the constraints is chosen to send the data to the sink. All nodes initially are assigned the same bandwidth ratio which makes constraints on other nodes which require higher bandwidth ratio. Furthermore, the transmission delay is not considered in the estimation of the end-to-end delay, which sometimes results in selecting routes that do not meet the required end-to-end delay. However, the problem of bandwidth assignment is solved in [15] by assigning a different bandwidth ratio for each type of traffic for each node.

SPEED [16] is another QoS based routing protocol that provides soft real-time end-to-end guarantees. Each sensor node maintains information about its neighbors and exploits geographic forwarding to find the paths. To ensure packet delivery within the required time limits, SPEED enables the application to compute the end-to-end delay by dividing the distance to the sink by the speed of packet delivery before making any admission decision. Furthermore, SPEED can provide congestion avoidance when the network is congested.

However, while SPEED has been compared with other protocols and it has showed less energy consumption than other protocols, this does not mean that SPEED is energy efficient, because the protocols used in the comparison are not energy aware. SPEED does not consider any energy metric in its routing protocol, which makes a question about its energy efficiency. Therefore to better study the energy efficiency of the SPEED protocol; it should be compared with energy aware routing protocols.

Felemban et al. [17] propose Multi-path and Multi-Speed Routing Protocol (MMSPEED) for probabilistic QoS guarantee in WSNs. Multiple QoS levels are provided in the timeliness domain by using different delivery speeds, while various requirements are supported by probabilistic multipath forwarding in the reliability domain.

Recently, X. Huang and Y. Fang have proposed multi constrained QoS multi-path routing (MCMP) protocol [18] that uses braided routes to deliver packets to the sink node according to certain QoS requirements expressed in terms of reliability and delay. The problem of the end-to-end delay is formulated as an optimization problem, and then an algorithm based on linear integer programming is applied to solve the problem. The protocol objective is to utilize the multiple paths to augment network performance with moderate energy cost. However, the protocol always routes the information over the path that includes minimum number of hops to satisfy the required QoS, which leads in some cases to more energy consumption. Authors in [19], have proposed the Energy constrained multi-path routing (ECMP) that extends the MCMP protocol by formulating the QoS routing problem as an energy optimization problem constrained by reliability, playback delay, and geo-spatial path selection constraints. The ECMP protocol trades between minimum number of hops and minimum energy by selecting the path that satisfies the QoS requirements and minimizes energy consumption.

Meeting QoS requirements in WSNs introduces certain overhead into routing protocols in terms of energy consumption, intensive computations, and significantly large storage. This overhead is unavoidable for those applications that need certain delay and bandwidth requirements. In our work, we combine different ideas from the previous protocols in order to optimally tackle the problem of QoS in sensor networks. In our proposal we try to satisfy the QoS requirements for real time applications with the minimum energy. Our QEMPAR routing protocol performs paths discovery using multiple criteria such as energy remaining, probability of packet sending, average probability of packet receiving and interference.

## 3. Proposed Protocol

In this section, we explain the assumptions and energy consumption model used in QEMPAR and describe the various constituent parts of the proposed protocol.

3.1 Assumptions

We assume that all nodes are randomly distributed in desired environment and each of them is assigned a unique ID. At start, the initial energy of nodes is considered equal. All nodes in the network are aware of their location (by GPS) and also are able to control their energy consumption. Because of this assumption has been that the nodes can communicate with other nodes outside their radio range in the absence of node in their radio transmission range.

Let us assume that nodes are aware of their remaining energy and also remaining energy of other nodes in their transmission radio range. We consider that each node can calculate its probabilities of packet sending and packet receiving with regard to link quality. Predications and decisions about path stability may be made by examining recent link quality information.

3.2 Energy Consumption Model

In QEMPAR, energy model is obtained from [20] that use both of the open space (energy dissipation $d^2$) and multi path (energy dissipation $d^4$) channels by taking amount the distance between the transmitter and receiver. So energy consumption for transmitting a packet of l bits in distance $d$ is given by Eq. (1).

$$E_{Tx}(l,d) = \begin{cases} lE_{elec} + l\varepsilon_{fs}d^2 & , d \leq d_0 \\ lE_{elec} + l\varepsilon_{mp}d^4 & , d > d_0 \end{cases} \quad (1)$$

In here $d_0$ is the distance threshold value which is obtained by Eq. (2), $E_{elec}$ is required energy for activating the electronic circuits. $\varepsilon_{fs}$ and $\varepsilon_{mp}$ are required energy for amplification of transmitted signals to transmit a one bit in open space and multi path models, respectively.

$$d_0 = \sqrt{\frac{\varepsilon_{fs}}{\varepsilon_{mp}}}. \quad (2)$$

Energy consumption to receive a packet of $l$ bits is calculated according to Eq. (3).

$$E_{Rx}(l) = lE_{elec}. \quad (3)$$

### 3.3 Link Suitability

The link suitability is used by the node to select the node at the next hop as a forwarder during the path discovery phase. Let $N_A$ be a set of neighbors of node A. Then our suitability function includes the PPS (Probability of Packet Sending), APPR (Average Probability of Packet Receiving) and $I_B$ (Interference of link A and B) and obtained by Eq. (4).

$$N\_H = \max_{B \in N_A} \{PPS_B + APPR_{N\_B} + 1/I_B + \frac{E_{r\_B}}{E_i}\}. \quad (4)$$

In here, $N\_H$ is the selected node at the next hop and $B$ is the node at the next hop. $PPS_B$ is the probability of packet sending of node $B$. Each node calculates the value of this parameter by Eq. (5). $APPR_{N\_B}$ is the average probability of packet receiving of all neighbors of node $B$ that obtained by Eq. (6). $I_B$ is interference of link between A and B. In this paper, $I_B$ is same signal to noise ratio (SNR) for the link between A and B.

$$PPS = \frac{Number-of-Successful-Sending-Packets}{Total-Number-of-Sending-Packets}. \quad (5)$$

$$APPR_{N\_B} = \sum_{j=1}^{N\_B} PPR_j. \quad (6)$$

The total merit (TM) for a path $p$ consists of a set of $K$ nodes is the sum of the individual link merit $l_{(AB)}$ along the path. Then the total merit is calculated by Eq. (7).

$$TM_p = \sum_{i=1}^{K-1} l_{(AB)_i}. \quad (7)$$

### 3.4 Paths Discovery Mechanism in QEMPAR

In multi-path routing, node-disjoint paths (i.e. have no common nodes except the source and the destination) are usually preferred because they utilize the most available network resources, and hence are the most fault-tolerant. If an intermediate node in a set of node-disjoint paths fails, only the path containing it node is affected, so there is a minimum impact to the diversity of the routes.

In first phase of path discovery procedure, each node collects the needed information about its neighbors by beacon exchange between them and then updates its neighboring table.

After this phase, each sensor node has enough information to compute the link suitability for its neighboring nodes.

### 3.5 Paths Assortment

After the execution of paths discovery phase and the paths have been constructed, we need to break a provided real time packet to few smaller packets, with sequence numbers assigned to each of them, in order to packet fast sending and consequently end to end delay decreasing. For this purpose, source node assortments the all paths according to hop counts of them in several classes. Then source node sends each tiny packet through separate paths. The tiny packet which its sequence number is 1 is sent through the path that has the least number of hops. Then other tiny packets with subsequent number according to the tiny packet number from packet number 2 to end through the paths with minimum hop count to maximum hop count. Because the sink to receive tiny packets consecutively. Fig.1 shows these operations.

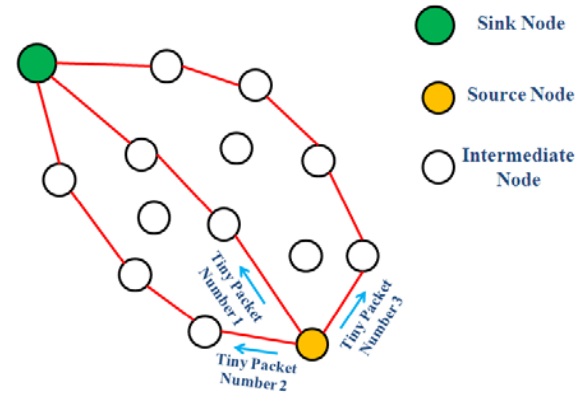

Fig. 1  Tiny packets sending through different paths

## 4. Simulation and Performance Evaluation

In this section, we present and discuss the simulation results for the performance study of QEMPAR protocol. We used GCC to implement and simulate QEMPAR and compare it with the MCMP protocol [18]. Simulation parameters are presented in Table 1 and obtained results are shown below.

The radio model used in the simulation was a duplex transceiver. The network stack of each node consists of IEEE 802.11 MAC layer with 40 meter transmission range.

We assume that the source node is located at (300, 300) meters.

Table 1: Simulation parameters

| Parameters | Value |
|---|---|
| Network area | 400 meters × 400 meters |
| Base station location | (0, 0)m |
| Number of sensors | 100 |
| Initial energy | 2J |
| $E_{elec}$ | 50 nJ/bit |
| $\varepsilon_{fs}$ | 10 pJ/bit/m$^2$ |
| $\varepsilon_{mp}$ | 0.0013 pJ/bit/m$^4$ |
| $d_0$ | 87 m |
| $E_{DA}$ | 5 nJ/bit/signal |
| Data packet size | 512 bytes |

We investigate the performance of the QEMPAR protocol in a multi-hop network topology. We study the impact of changing the packet arrival rate on end-to-end delay, packet delivery ratio, and energy consumption. We change the real-time packet arrival rate at the source node from 5 to 50 packets/sec.

### 4.1 Average End-to-End Delay

The average end-to-end delay is the time required to transfer data successfully from source node to the destination node.

Fig. 2 shows the average end to end delay for QEMPAR and MCMP. In this evaluation, we change the packet arrival rate at the source node, and measure the delay.

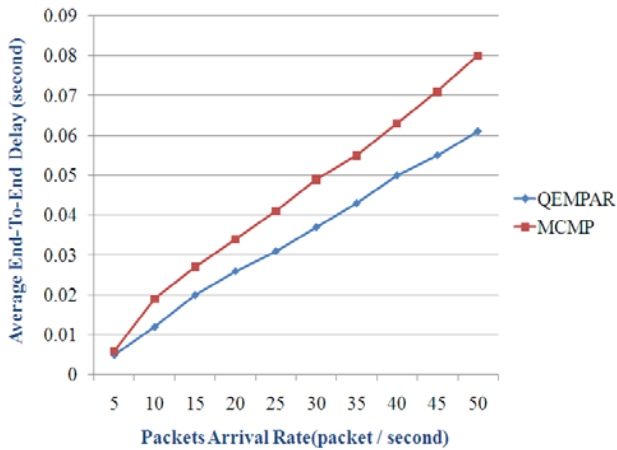

Fig. 2 Average end to end delay

As it can be seen, proposed protocol has performance better than MCMP in average end to end delay.

### 4.2 Average Energy Consumption

The average energy consumption is the average of the energy consumed by the nodes participating in message transfer from source node to the destination node.

Fig. 3 shows the results for energy consumption in two protocols. As it can be seen, in our protocol, energy consumption for packet sending is some deal optimize in comparison to the MCMP.

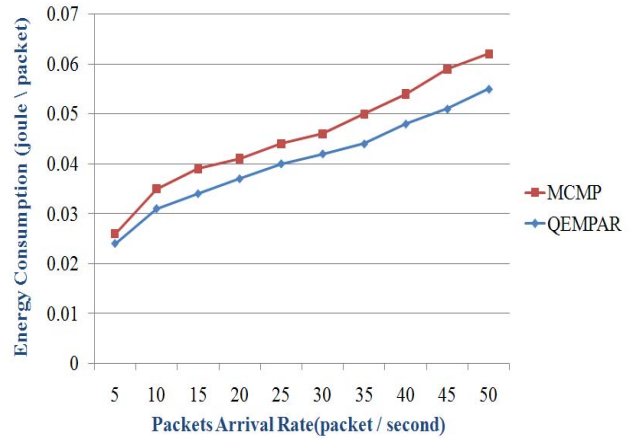

Fig. 3 Average energy consumption

## 5. Conclusion

In this paper, we propose the new multi path routing algorithm for real time applications in wireless sensor network namely QEMPAR which is QoS aware and can increase the network lifetime. Our protocol uses four main metrics of QoS with special relation in path discovery mechanism. Simulation Result shows that the performance of QEMPAR in end to end delay is optimized compared to the MCMP protocol.